# Home Environment and Students' Creative Thinking: An Educational Data Science Analysis of PISA 2022

George X. Wang, New York University, USA

Yuyang Shen, University of North Texas, USA

# Introduction

This study investigates how students' exposure to resources in their home environments relates to creative thinking performance, using data from the PISA 2022 Creative Thinking assessment. It focuses on two primary questions: (1) How strongly is exposure to cultural, educational, and digital resources associated with creativity? (2) Do students perform better on divergent thinking tasks when physically engaged or digitally stimulated? Drawing on a sample of 15,425 students from 60 countries, the study applies high-dimensional regression and factor analysis to identify patterns across a wide range of exposure variables. The goal is to offer evidence-based insights for educators, families, and policy-makers seeking to foster creativity outside formal classroom contexts.

# Background

We ground our inquiry in prevailing creativity theory emphasizing the role of the environment (or "press") in creative development. Creativity is commonly defined as **"the interaction among aptitude, process, and environment by which an individual or group produces a perceptible product that is both novel and useful as defined within a social context"**. This **system's perspective** (Plucker et al., 2004) explicitly includes the environment alongside individual ability and cognitive process in the creative act. In educational contexts, the environment encompasses family, peers, and resources that can press for or against creativity (Thornhill-Miller et al., 2023). We draw on Rhodes' 4P model, particularly the *Press* dimension, which highlights how physical and social surroundings can nurture or hinder students' creativity (Rhodes, 1961). A supportive home environment, such as one rich in books, artistic materials, or encouragement for original thinking, may function as a positive "press" that helps students "think outside the box."

In the digital age, our framework also incorporates emerging conceptions of digital creativity support. We adapt the notion of "digital exposure" as engagement with digital tools that support creative ideation, production, or reflection, as recently proposed by Ceh et al. (2024) in their study of how technology mediates creativity. Digital creativity support frameworks emphasize features like tools' functions, content, and community that aid different stages of the creative process (Ceh et al., 2024). Guided by this, we conceptualize students' digital exposure in terms of access to and use of creative digital resources (such as computers, internet connectivity, and creative software or platforms) at home. Meanwhile, physical exposure refers to tangible cultural and learning resources in the home: books, musical instruments, art supplies, etc. These can stimulate imaginative thinking and creative practice.

Our analysis also leverages perspectives from educational assessment and learning environment research. Large-scale assessments like PISA have begun to measure "non-cognitive" skills and creative competencies, providing new data to explore environmental influences (Barbot & Kaufman, 2025). We align with an ecological systems view in positing that creative development is situated: the home micro-environment, such as family attitudes, resources available, learning activities, can significantly shape a student's opportunity to develop creativity (VYGOTSKY, 1967). Prior studies suggest, for example, that students with supportive family attitudes towards creativity or abundant learning materials often show higher creative engagement (Hopson et al., 2024). Moreover, creativity and academic achievement, while related, do not completely overlap – PISA 2022 results showed that about half of students who excelled in creative thinking were not top performers in academic domains (Earp, 2024). This underscores creativity as a distinct competency potentially nurtured by different factors than those fostering traditional academic success.

# Methodology

## Sample and Data Sources

This study uses publicly available data from the Programme for International Student Assessment (PISA) 2022, specifically the Creative Thinking assessment and its corresponding student background questionnaire. The analytic sample includes 15,425 students from 60 participating countries and economies that administered the creative thinking option.

The dataset combines two main components:

1. **Creative Thinking Cognitive Assessment**: This assessment was introduced for the first time in the 2022 PISA cycle. It includes open-ended tasks designed to assess students' ability to generate, evaluate, and improve ideas across four domains: written expression, visual expression, social problem solving, and scientific problem solving.
2. **Student Background Questionnaire**: This self-report survey collects extensive information on students' home environments, learning habits, digital access, attitudes, and family background. Over one thousand variables are available for analysis, allowing for broad exploration of environmental influences.

## Research Design

This study employs a secondary quantitative analysis with a correlational design. The primary goal is to examine the relationship between home environmental factors and creative thinking performance using large-scale cross-national data. Data cleaning and merging were performed using Python and R. We selected a subset of variables relevant to home environment and student-level experience, guided by theory and prior literature. These included indicators of physical resources (e.g., number of books, musical instruments) and digital resources (e.g., number of screen devices, internet quality, digital activity frequency). Variables with excessive missingness or insufficient variance were excluded from further analysis. Descriptive statistics and missing data patterns were examined before analysis. Student weights provided by OECD were applied to adjust for complex sampling design and ensure population-level inference.

**Exploratory Analysis**

We conducted an initial exploratory analysis to understand the structure of the data. We first examined the distribution of the creative thinking scores. The score distribution was approximately normal (unimodal with modest skew), supporting the use of parametric techniques in subsequent analyses. We also computed zero-order correlations and created visualizations (e.g. score histograms and boxplots) to observe how creativity scores varied across different levels of key home variables. For instance, a boxplot of creativity scores by number of books at home suggested a positive trend – students reporting more books tended to achieve higher creative thinking scores on average. Similarly, creative scores appeared higher for students with greater access to digital devices at home, hinting at the relevance of digital exposure. These exploratory patterns motivated a more formal multivariate analysis. The exploratory phase was used to identify potential patterns in the data and inform later model construction.

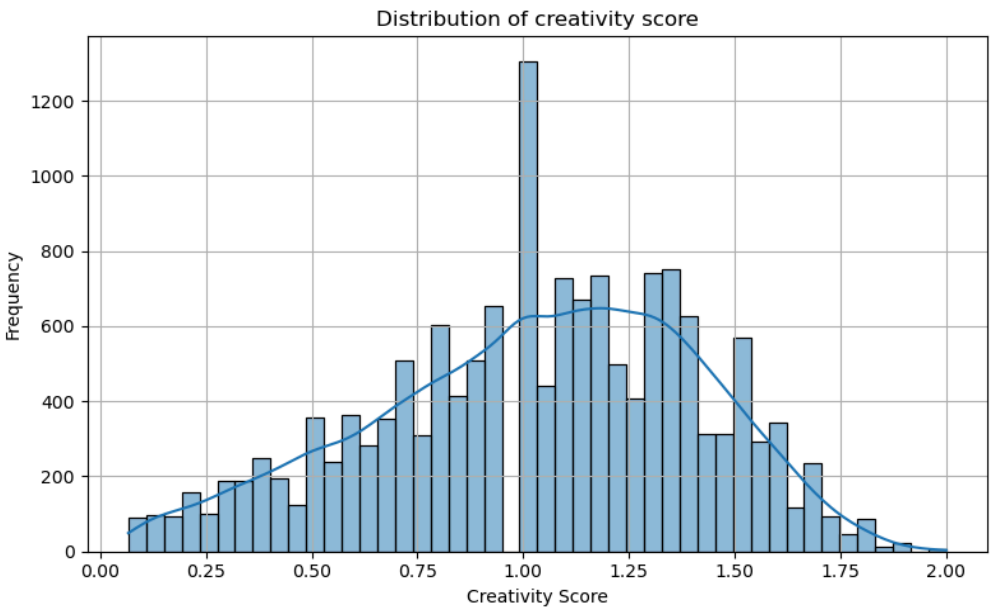

*Table 1.* Distribution of Student Creativity Scores from the PISA 2022 Assessment

## Confirmatory Factor Analysis (CFA)

To model the latent structure of home environment variables, we conducted a Confirmatory Factor Analysis. We proposed a two-factor structure:

- **Physical Exposure**, representing tangible cultural and educational resources in the home environment.
- **Digital Exposure**, representing access to and engagement with technology and digital learning tools.

Each factor was constructed using multiple observed indicators selected from the student questionnaire. The CFA was conducted using maximum likelihood estimation with robust standard errors. Model fit was assessed using standard indices (CFI, TLI, RMSEA), and standardized loadings were used to confirm the validity of the factor structure. Factor scores derived from the CFA were used as composite predictors in subsequent modeling steps.

| Exposure Type | Question | Variable | Number of Participants | Scale |
|---|---|---|---|---|
| Digital Exposure | How many <digital devices> with screens are there in your <home>? | ST253Q01JA | 125180 | 8 |
| Digital Exposure | How many of the following <digital devices> are in your <home>? (Laptop computers or notebooks) | ST254Q03JA | 120965 | 5 |
| Digital Exposure | How many of these items are there at your <home>? (Works of art (e.g. paintings, sculptures, <country-specific example>) | ST251Q07JA | 122314 | 4 |
| Digital Exposure | Which of the following are in your <home>? (A computer (laptop, desktop, or tablet) that you can use for school work) | ST250Q02JA | 124191 | 2 |
| Physical Exposure | How many of the following types of books are in your <home>? (Contemporary literature) | ST256Q03JA | 114340 | 5 |
| Physical Exposure | How many of these items are there at your <home>? (Works of art (e.g. paintings, sculptures, <country-specific example>) | ST251Q07JA | 122314 | 4 |

*Table 1. Selected Indicators of Digital and Physical Exposure in the Home Environment*

# Results

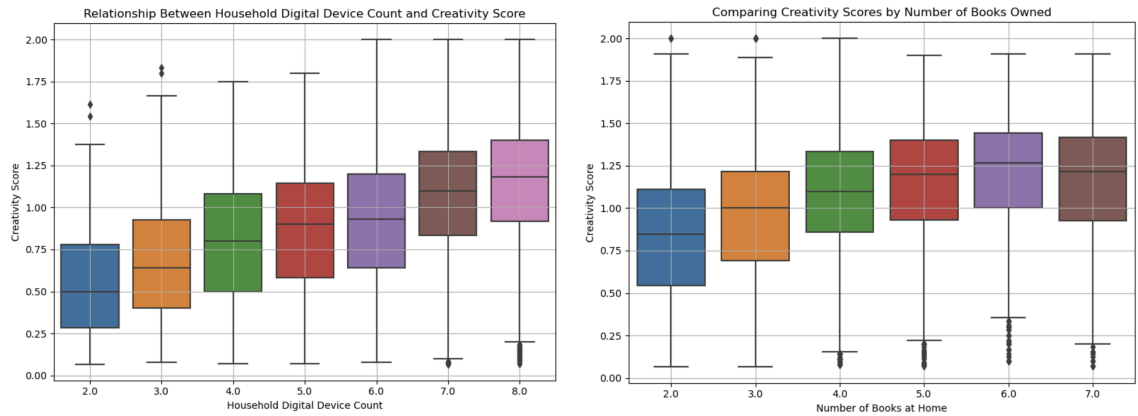

*Figure 2 and Figure 3. Associations Between Creativity Scores and Home Resources: Digital Devices (Left) and Number of Books (Right)*

Figures 2 and 3 present boxplots illustrating the association between key home resources and students' creative thinking scores. The boxplot on the left shows the relationship between household digital device count and creativity score. There is a clear positive trend: students with access to only two digital devices exhibit median creativity scores below 0.75, while those with access to eight devices reach median scores near 1.25. The steady upward progression in median and interquartile range demonstrates that greater access to digital technology corresponds with higher creative performance. The boxplot on the right displays creativity scores by number of books owned at home. Students reporting two books have a median creativity score just above 0.75, while those with seven books reach a median closer to 1.3. The shift in the interquartile range across book categories suggests a similarly strong positive association between the abundance of physical reading materials and creative thinking achievement. Together, these figures provide visual evidence that both digital and physical resources in the home environment are positively linked to students' creative performance.

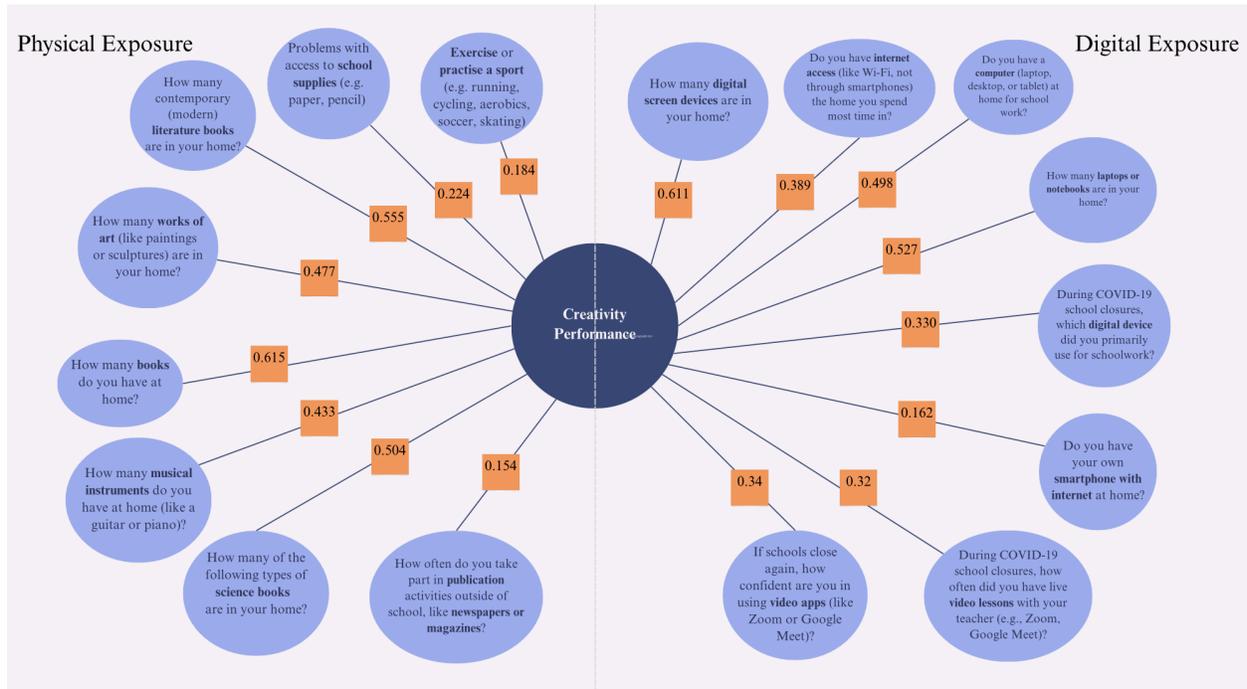

*Figure 4. Relationships between select home environment variables and creative thinking performance, grouped into "Physical Exposure" (left) and "Digital Exposure" (right). Numbers in orange boxes indicate the strength of association (standardized coefficient of correlation) between each observed variable (blue) and students' creative performance. Key resources like number of books at home (loading ≈0.62) and number of digital devices (≈0.61) show among the strongest positive links with creative thinking outcomes.*

To formally capture the structure underlying these resource variables, a confirmatory factor analysis (CFA) was conducted using items from the student questionnaire. The analysis specified two latent factors: Physical Exposure and Digital Exposure. Each are measured by multiple observed indicators, as shown in Figure 4. The model demonstrated excellent fit, with a Comparative Fit Index (CFI) of 0.971 and a Root Mean Square Error of Approximation (RMSEA) of 0.038. All indicators loaded significantly onto their respective factors, with standardized loadings generally ranging from 0.4 to 0.7. For example, the variable "How many books do you have at home?" loaded at 0.615 on the Physical Exposure factor, while "How many digital screen devices are in your home?" loaded at 0.611 on the Digital Exposure factor. Other notable contributors to the Physical Exposure construct included the number of works of art at home (0.477), access to musical instruments (0.504), and access to science books (0.433). Within the Digital Exposure construct, high-loading variables included the number of laptops or notebooks (0.527), internet access at home (0.498), and the availability of a computer for schoolwork (0.389). Variables reflecting confidence in using video apps (0.34) and frequency of live video lessons during COVID-19 school closures (0.32) also contributed to the digital factor. The two latent factors were moderately correlated (r = 0.47), indicating that while homes with abundant physical resources often also possess digital resources, these constructs remain sufficiently distinct to warrant separate interpretation.

When both factors were entered together in the regression, physical and digital exposures each contributed unique explanatory power. There is no indication that one simply proxies the other; rather, they appear to be complementary dimensions of a creative home environment. The Physical Exposure index had a slightly larger standardized effect than Digital Exposure in most models, but both were statistically significant. Together, these home environment factors explained a meaningful portion of variance in scores.

## Discussion

The present study advances the field of creativity research by providing large-scale, quantitative evidence on the role of both physical and digital home environments in shaping adolescent creative thinking performance. Drawing on the theoretical foundation that creativity emerges from the dynamic interaction among individual aptitudes, cognitive processes, and environmental influences (Plucker et al., 2004), our analysis offers several contributions that clarify and extend existing scholarship.

First, the findings substantiate long-standing environmental models of creativity, such as Rhodes' "4P" framework and the "press" component articulated in the systems theory of creativity (Rhodes, 1961; Plucker et al., 2004). Our use of PISA 2022 data from over 15,000 students across 60 countries provides unprecedented empirical support for the measurable impact of environmental resources at a global level. Notably, the strong associations observed between home resources, especially the number of books and digital devices, and creative performance reinforce the view that creativity is not simply an innate trait, but one that is nurtured and developed in context-rich environments (Sternberg & Lubart, 1998). The distinct predictive value of both physical and digital exposures refines the understanding of environmental "press" for the 21st century, suggesting that material, cultural, and technological supports all play a role in cultivating creative potential.

Second, this study bridges the gap between traditional creativity research and emerging work on digital creativity. While access to books and cultural items has long been recognized as a foundation for creative growth (Kim, 2011), the present findings add empirical weight to the argument that digital resources and technology-supported experiences are equally relevant in contemporary creative development (Ceh et al., 2024). The positive associations between creative performance and variables such as number of digital devices, confidence using video platforms, and internet access align with current research on "creativity support tools" (Shneiderman, 2007; Lee et al., 2015). These results support an increasingly nuanced perspective that, when harnessed for production, collaboration, or reflection, digital technologies enrich rather than constrain students' creative abilities. Such findings contribute to ongoing scholarly debates about whether digital engagement is beneficial, detrimental, or context-dependent for creative thinking (Kaufman & Beghetto, 2013).

The limitations of the present study must be acknowledged. The cross-sectional and correlational design precludes strong causal inference, and further experimental or longitudinal research is needed to determine the direct effects of enhancing specific home resources on creativity outcomes. Additionally, the reliance on self-reported measures introduces the possibility of reporting biases. The generalizability of findings to different cultural contexts, while strengthened by the international sample, may still be moderated by local values, norms, and educational systems; future research could explore cultural variability in greater depth(Lewin et al., 2016).

Future research directions include examining the interaction effects between physical and digital resources. Specifically, how digital exposure can compensate for limited physical resources, or vice versa. Qualitative investigations could shed light on how students use books or digital devices creatively in daily life, and whether certain types of engagement (such as collaborative digital projects or reading diverse genres) are especially potent. Finally, policy-oriented studies could assess the impact of targeted interventions, such as book distribution or technology loan programs, on creativity development over time.

## Conclusion

This study offers compelling international evidence that both physical and digital resources in the home environment play significant, independent, and complementary roles in shaping adolescents' creative thinking abilities. Drawing on extensive data from the PISA 2022 Creative Thinking assessment, the analysis demonstrates that access to books, artistic materials, and digital technologies contributes meaningfully to students' creative performance across diverse national settings. By systematically applying advanced educational data science methods to a large-scale, multi-country dataset, this research validates and extends foundational theories of creativity that emphasize the interaction between individual aptitude and environmental context. The clear associations observed between home resources and creative outcomes reinforce the importance of considering both traditional and emerging forms of exposure in educational planning and policy.

These findings have direct implications for efforts to promote creativity and equity in education. Policies that prioritize expanding access to learning-rich environments have the potential to foster creative potential in learners from all backgrounds. As creativity becomes increasingly vital in the modern world, educators and policymakers should view the enrichment of students' material, cultural, and technological environments as a core strategy for cultivating the creative thinkers of tomorrow. In sum, this work illustrates the powerful synergy between educational data science and creativity research. By bridging large-scale empirical analysis and theory-driven inquiry, it advances understanding of how environments nurture creative development and provides actionable guidance for future educational innovation.